\documentclass[12pt]{article}
\usepackage{amsmath}
\usepackage{epsfig}
\usepackage{amssymb}
\usepackage{hyperref}
\usepackage{bbold}
\usepackage{braket}
\usepackage{cite}

\input{epsf}
\def\be{\begin{equation}}
\def\ee{\end{equation}}
\def\beq{\begin{equation}}
\def\eeq{\end{equation}}
\def\beqa{\begin{eqnarray}}
\def\eeqa{\end{eqnarray}}
\def\ba{\begin{eqnarray}}
\def\ea{\end{eqnarray}}
\def\bea{\begin{eqnarray}}
\def\eea{\end{eqnarray}}

\def\beq{\begin{equation}}
\def\eeq{\end{equation}}
\def\beeq{\begin{eqnarray}}
\def\eeeq{\end{eqnarray}}
\def\to{\rightarrow}
\def\nn{\nonumber}

\def\b0{b_0}

\def\b0{b_0}

\begin{document}

\begin{titlepage}
\renewcommand{\thefootnote}{\fnsymbol{footnote}}
\begin{flushright}
TIF-UNIMI-2024-01
     \end{flushright}
\par \vspace{10mm}
\begin{center}
{\large \bf
Higgs production at RHIC and the positivity\\[2mm] of the gluon helicity distribution}\\

\vspace{8mm}

\today
\end{center}

\par \vspace{2mm}
\begin{center}
{\bf Daniel de Florian${}^{\,a}$,}
\hskip .2cm
{\bf Stefano Forte${}^{\,b}$,}
\hskip .2cm
{\bf Werner Vogelsang${}^{\,c}$}\\[2mm]
\vspace{5mm}
${}^{a}$ International Center for Advanced Studies (ICAS) and ICIFI, \\ ECyT-UNSAM,
Campus Miguelete, \\ 25 de Mayo y Francia, (1650) Buenos Aires, Argentina\\[2mm]
${}^{b}$ Tif Lab, Dipartimento di Fisica, Universit\`{a} di Milano and INFN,\\
Sezione di Milano, Via Celoria 16, I-20133 Milano, Italy \\[2mm]
${}^{c}$ Institute for Theoretical Physics, T\"ubingen University, 
Auf der Morgenstelle 14, \\ 72076 T\"ubingen, Germany\\[2mm]
\end{center}


\vspace{9mm}
\begin{center} {\large \bf Abstract} \end{center}
We show that the negative polarized gluon
distribution $\Delta g$ found in a recent global next-to-leading order QCD analysis of the nucleon helicity structure is incompatible with the fundamental requirement that physical cross-sections 
must not be negative. Specifically, we show that the fact that this
polarized gluon strongly violates the positivity condition $|\Delta
g|\leq g$ in terms of the unpolarized gluon distribution $g$ leads to
negative cross-sections for Higgs boson production at RHIC as a
physical process,  
implying  that this negative $\Delta g$ 
is unphysical. 

\vspace*{1cm}
\begin{center}
{\Large
{\it Dedicated to the memory of Stefano Catani}}
\end{center}
\vspace*{2cm}
\end{titlepage}  

\renewcommand{\thefootnote}{\fnsymbol{footnote}}


Understanding the quark and gluon spin structure of the proton is a
key focus of modern nuclear and particle physics. 
An important component of this endeavor is the precise determination
of the proton helicity parton distribution functions (PDFs). 
The gluon helicity PDF $\Delta g(x,\mu)$, in particular, has received
much attention in this context as its integral over 
all momentum 
fractions $x$ measures the gluon spin contribution to the proton spin
and hence could  
hold the key to decomposing the proton spin into its partonic
contributions. A celebrated discovery  
was made in 2014, when it was
shown~\cite{deFlorian:2014yva,Nocera:2014gqa} that data from the  
Relativistic Heavy Ion Collider (RHIC)~\cite{RHICSPIN:2023zxx} provided
evidence for a non-vanishing and positive $\Delta g$ in the region
$0.05\lesssim x\lesssim 0.2$. This 
finding was obtained on the basis of a global next-to-leading order
(NLO) QCD analysis of the 
world data on polarized (semi-)inclusive deep-inelastic scattering and
polarized $pp$ scattering.  
It was confirmed in additional
studies~\cite{Sato:2016tuz,DeFlorian:2019xxt} and later substantially 
corroborated when further sets of RHIC data became
available~\cite{RHICSPIN:2023zxx}.  

The need for a  positive $\Delta g$ was recently called into question
in Refs.~{\cite{Zhou:2022wzm,Cocuzza:2022jye,Karpie:2023nyg}. 
Again in the context of an NLO analysis (in the
$\overline{\mathrm{MS}}$ scheme) the authors found PDFs featuring a
negative 
gluon helicity PDF, $\Delta g<0$, thereby suggesting that negative
gluon polarization is also possible. These PDFs are delivered as an
ensemble of replicas, quantifying the PDF
uncertainty. Figure~\ref{fig1} shows the 78 PDF replicas  with
negative $\Delta g$ from  Ref.~\cite{Cocuzza:2022jye}, 
available at~\cite{jamgithub} in LHAPDF~\cite{Buckley:2014ana}
format. We show results at factorization scales $\mu=\sqrt{10}$~GeV
(left) and $\mu=125$~GeV (right).  

A striking feature of the PDFs with negative $\Delta g$ proposed
in~{\cite{Zhou:2022wzm,Cocuzza:2022jye,Karpie:2023nyg}   
is that they strongly violate the positivity condition
\beq\label{eq1}
|\Delta g(x,\mu)|\,\leq g(x,\mu)
\eeq
at momentum fractions $x\gtrsim 0.25$. Indeed, as the authors state,
the negative $\Delta g$ PDFs are 
obtained only when the positivity condition is relaxed in the
analysis. The violation of the inequality~(\ref{eq1}) 
is evident from Fig.~\ref{fig1} where we also show in both panels the
replicas for the corresponding unpolarized gluon 
distribution as obtained in the same analysis~\cite{Cocuzza:2022jye}. 

\begin{figure}[h!]
\vspace*{4mm}
\hspace*{-1.4cm}
\epsfig{figure=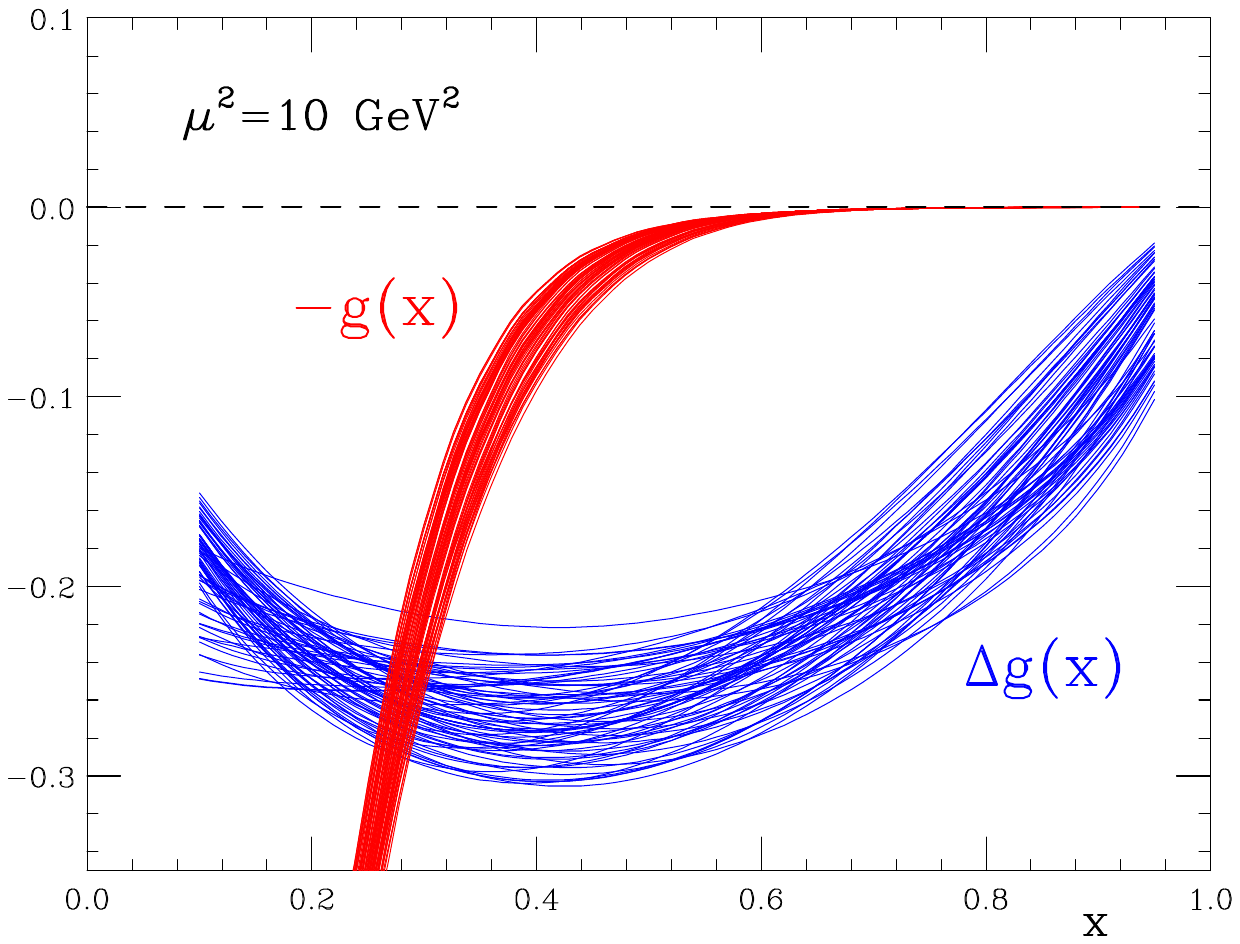,width=0.6\textwidth,clip=}
\hspace*{-1.6cm}
\epsfig{figure=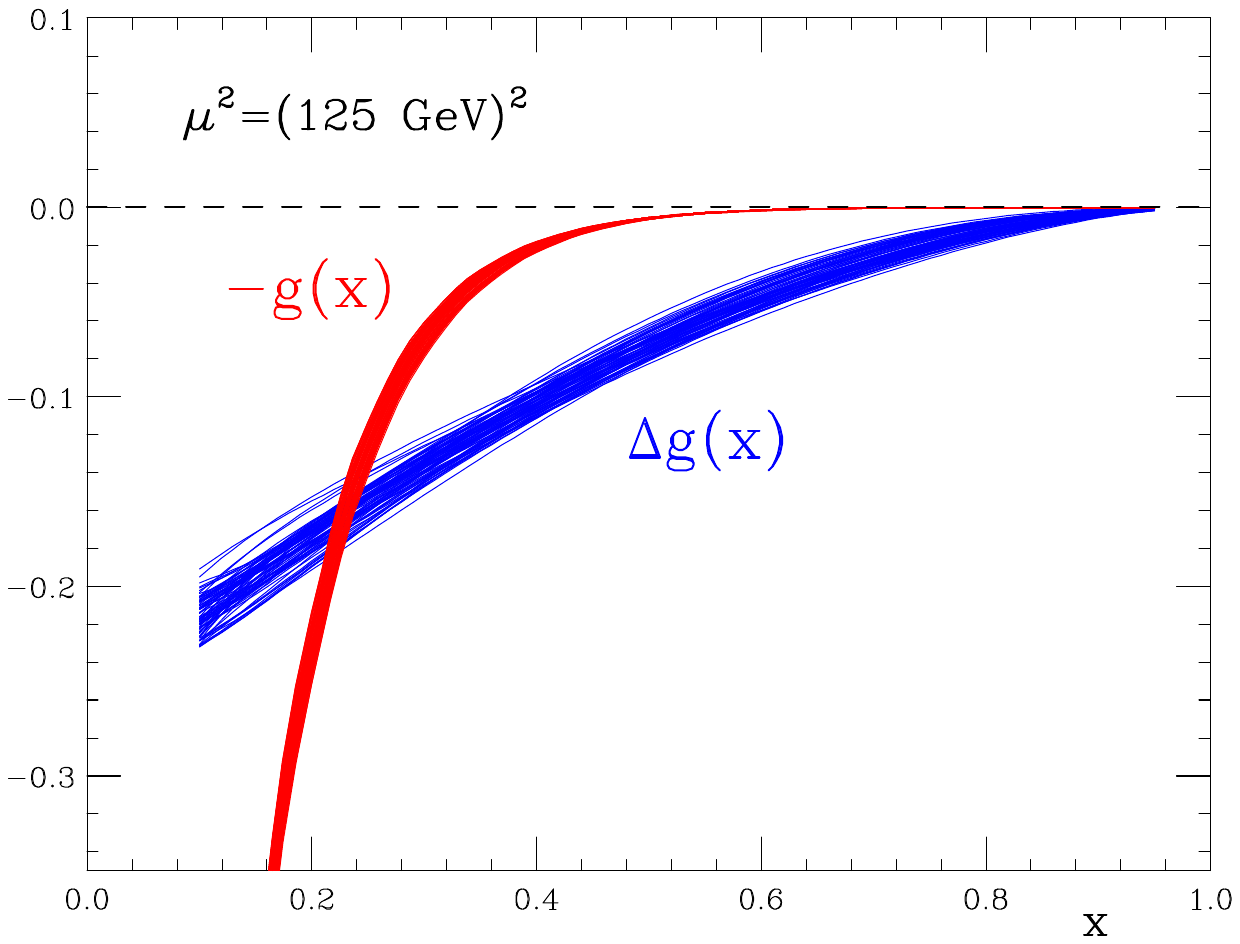,width=0.6\textwidth,clip=}
\vspace*{-1.3cm}
\caption{ \label{fig1} {\it PDF replicas of Ref.~\cite{Cocuzza:2022jye} for $-g(x,\mu)$ and $\Delta g(x,\mu)$ at $\mu=\sqrt{10}$~GeV (left) and
$\mu=125$~GeV (right).}}
\end{figure}                 

Condition~(\ref{eq1}) arises of course from the fact that $g=g_++g_-$
and $\Delta g =g_+-g_-$ where $g_+,g_-$ 
are the distributions for gluons with positive or negative helicity
inside a proton with positive helicity, respectively. At leading order
(LO) in perturbative QCD these can be regarded as number densities and
hence positive\footnote{Here and elsewhere we use the word positive to
mean non-negative.},  so 
Eq.~(\ref{eq1}) strictly holds. This is a consequence of the fact that
there exist physical processes for which at leading order the
physically measurable cross-section, which is a probability and thus
positive, is proportional to the PDFs. However, beyond LO the
cross-section is obtained by convoluting the PDF with a partonic
cross-section  (coefficient function). The cross-section remains of
course positive, but now the coefficient function and the PDF depend
on the factorization  
scheme and hence are not necessarily separately positive, so strict
positivity of $g_+,g_-$ does not need to hold any
longer~\cite{Altarelli:1998gn,Forte:1998kd,Candido:2020yat,Candido:2023ujx,Collins:2021vke}. As  
a result, solutions  
with $|\Delta g|\geq g$ are in principle formally possible. 

Such violations
of positivity however must have the size of the higher-order
corrections, because the possible violation of positivity of PDFs must
be compensated by the higher-order corrections to the coefficient
functions so that the physical cross-section remains positive. Indeed,
In Ref.~\cite{Altarelli:1998gn} next-to-leading order (NLO) positivity
bounds on polarized PDFs in the $\overline{\rm MS}$ scheme were
derived by requiring positivity of NLO cross-sections. These NLO
bounds were used to derive a bound on the polarized gluon distribution
in $x$ space~\cite{Forte:1998kd}, which was compared to the naive LO
bound Eq.~(\ref{eq1}) and found to differ from it at the percent level
except at very small $x\lesssim 10^{-3}$. 
In contrast to this, the violation
of positivity of the PDFs from Refs.~{\cite{Zhou:2022wzm,Cocuzza:2022jye,Karpie:2023nyg} exhibited in Fig.~\ref{fig1} is 
much larger: in fact,  $|\Delta g|$ exceeds $g$ by a large
factor. This suggests that these PDFs may lead to unphysical
predictions.  

To see how this may happen, we recall how positivity bounds can be
derived at any perturbative order~\cite{Altarelli:1998gn}. The
derivation is based on the observation that  
physically observable cross-sections -- 
and theoretical predictions thereof -- are proportional to the number
of observed events, and thus cannot be negative. Using the
spin-dependent cross-section 
for some reaction in polarized $pp$ scattering as an example, we must have
\beq\label{eq2}
|A_{\mathrm{LL}}|\,\leq\,1\,,
\eeq
where
\beq\label{eq3}
A_{\mathrm{LL}}\,\equiv\,\frac{\Delta\sigma}{\sigma}\,\equiv\,\frac{\sigma_{++}-\sigma_{+-}}{\sigma_{++}+\sigma_{+-}}\,,
\eeq
with $\sigma_{++}$ ($\sigma_{+-}$) the cross-section when the two
colliding protons have the same (or opposite) 
helicities. (For simplicity, we are considering a parity-conserving
interaction).  
The condition~(\ref{eq2}) must apply
to any physical cross-section, regardless of whether it has been
measured, or even whether it is practically measured or measurable in
an actual experiment. Positivity bounds on any polarized PDF at, say
NLO can then be derived by imposing the condition Eq.~(\ref{eq2}) on a
set of suitably chosen pairs of polarized and unpolarized NLO
cross-sections. For instance, the  
positivity bounds of Ref.~\cite{Altarelli:1998gn} were obtained by
imposing the condition on polarized and unpolarized deep-inelastic
scattering, as well as for Higgs production in gluon--proton
scattering.

Because the bound Eq.~(\ref{eq2}) implies a bound on the polarized
PDFs, it follows that PDFs that violate this positivity bound lead to
unphysical negative cross-sections. 
Given the enormous violation of positivity for the gluon PDF as seen
in Fig.~\ref{fig1}, one may immediately 
ask whether there could be a physical observable for which an NLO
prediction based on this gluon density 
violates the condition Eq.~(\ref{eq2}). We will now show that this is
indeed the case, rendering the solutions  
of~{\cite{Zhou:2022wzm,Cocuzza:2022jye,Karpie:2023nyg} with negative
  $\Delta g$ unphysical.  
A suitable candidate for this purpose is an observable that is
gluon-driven at tree level, and probes the region $x\gtrsim 0.25$ of
momentum fractions 
where the gluon distribution in Fig.~\ref{fig1} violates positivity,
such as the Higgs production 
cross-section in $pp$ scattering. A dominant contribution to this
process is gluon-gluon fusion, 
$gg\to H$, through a top quark loop. To lowest order in QCD no
additional partonic channels involving incoming quarks contribute to
this process.  

This process is of course of paramount importance at the LHC, and its
total cross-section has been studied in this 
context in great theoretical detail (see
Refs.~\cite{LHCHiggsCrossSectionWorkingGroup:2011wcg,
  Dittmaier:2012vm,LHCHiggsCrossSectionWorkingGroup:2013rie,LHCHiggsCrossSectionWorkingGroup:2016ypw}).  
Here we will instead consider Higgs production at RHIC energy,
$\sqrt{S}=510$~GeV, using 
a range of Higgs masses $m_H$ between 100~GeV and 250~GeV. Although
this is not relevant 
for canonical Higgs phenomenology, it will allow us to access
a perturbative cross-section in the kinematic regime where $x\gtrsim
0.25$, since the  
lowest $x$-value
probed in the PDFs at leading order for a given Higgs mass is
$m_H^2/S$, with $\sqrt{S}$ the $pp$ center-of-mass energy.

The unpolarized and spin-dependent cross-sections for $pp\to H X$ may,
up to power corrections, be written in factorized 
form as 
\beeq\label{eq4}
\sigma^{pp\to H}&=&\sigma_0 \sum_{i,j} \int_\tau^1 dx_1\,\int_{\tau/x_1}^1 dx_2\,
f_i(x_1,\mu)\, f_j(x_2,\mu)\,\omega^{ij\to
  H}\left(z=\frac{\tau}{x_1x_2},\alpha_s(\mu),\frac{\mu}{m_H}\right)\,,\nn\\[2mm]  
\Delta \sigma^{pp\to H}&=&\sigma_0 \sum_{i,j} \int_\tau^1
dx_1\,\int_{\tau/x_1}^1 dx_2\, 
\Delta f_i(x_1,\mu)\,\Delta f_j(x_2,\mu)\,\Delta\omega^{ij\to H}\left(z=\frac{\tau}{x_1x_2},\alpha_s(\mu),\frac{\mu}{m_H}\right),\;\;\;
\eeeq
where $\tau=m_H^2/S$ and $\omega^{ij\to H},\Delta\omega^{ij\to H}$ are
normalized hard-scattering functions that are computed in perturbation
theory.  
They are defined as in Eq.~(\ref{eq3}) by $\omega^{ij\to
  H}\equiv\frac{1}{2}(\omega^{ij\to H}_{++}+\omega^{ij\to H}_{+-})$ 
and $\Delta \omega^{ij\to H}\equiv\frac{1}{2}(\omega^{ij\to
  H}_{++}-\omega^{ij\to H}_{+-})$, where $\omega^{ij\to
  H}_{\lambda_i\lambda_j}$ is the  
cross section for incoming partons $i,j$ with helicities
$\lambda_i,\lambda_j$. The normalization $\sigma_0$ is the same 
for $\sigma^{pp\to H}$ and $\Delta \sigma^{pp\to H}$ and is given by
\beq
\sigma_0\,=\,\frac{\alpha_s^2 |A|^2}{256 \pi v^2}\,,
\eeq
with $\alpha_s$ the strong coupling and $v=246$~GeV the Higgs vacuum
expectation value. The factor $|A|^2$  
results from the coupling of the two gluons to the Higgs boson via a
heavy-quark loop. Ignoring contributions from  
charm and bottom quarks and keeping only the top quark of mass $m_t$,
one has, at lowest order (see~\cite{Djouadi:1991tka,Dawson:1990zj}): 
\beq 
A\,=\,\tau_q \left(1+(1-\tau_q)\arcsin^2(1/\sqrt{\tau_q})\right)\,,
\eeq
where $\tau_q\equiv 4m_t^2 /m_H^2$, and beyond leading order the
hard-scattering functions also depend on $m_t$. This expression was
originally obtained for the spin-averaged cross-section, but we 
have checked that it also holds for the spin-dependent one. One may
further assume 
that the top quark is infinitely heavy. In the effective theory
defined by this assumption, one has $|A|^2=4/9$.  
In any case, the factor $|A|^2$ cancels in the spin asymmetry. 
For simplicity, we have chosen the factorization and renormalization
scales to be the same in Eq.~(\ref{eq4}) 
and denoted them by $\mu$. For our numerical results further below, we
will set $\mu=m_H/2$, a value that is known to lead to faster
convergence of  the perturbative expansion of the Higgs
cross-section~\cite{Anastasiou:2016cez}. However, none of our results
depends 
qualitatively on this choice.

As mentioned, to lowest order, $gg\to H$ is the only contributing
channel: because this is  a $2\to 1$ reaction, it is characterized  
by $\hat{s}=m_H^2$, corresponding to $z=1$, where $\hat{s}=x_1x_2S$ is
the partonic center-of-mass energy squared.  
Correspondingly, the partonic hard-scattering function is given by a
Dirac delta  $\delta(1-z)$ at this order. The NLO  
corrections to the partonic hard-scattering functions have been computed in the 
$\overline{\mathrm{MS}}$ scheme in Refs.~\cite{Djouadi:1991tka,Dawson:1990zj}
for the spin-averaged and in~\cite{Altarelli:1998gn} for the
spin-dependent cross-section. For the $gg$-channel 
we have, up to corrections of order $\alpha_s^2$:
\beeq
\omega^{gg\to H}\left(z,\alpha_s,r\right)&=&\delta(1-z)\,+\,\frac{\alpha_s}{\pi}
\left\{ \delta(1-z) \left(\frac{11}{2}+\pi^2\right)-\frac{11}{2}
(1-z)^3\right. \nn\\[2mm] 
&+&\left.6\left( 1-z+z^2\right)^2\left[2
  \left(\frac{\ln(1-z)}{1-z}\right)_+  -\frac{\ln(z)}{1-z}- 
\frac{\ln(r^2)}{(1-z)_+}
\right]\right\}\,,\nn\\[2mm]
\Delta\omega^{gg\to
  H}\left(z,\alpha_s,r\right)&=&\delta(1-z)\,+\,\frac{\alpha_s}{\pi} 
\left\{ \delta(1-z) \left(\frac{11}{2}+\pi^2\right)+\frac{11}{2}
(1-z)^3\right. \nn\\[2mm] 
&+&\left.6 z \left (2-3z+2z^2\right)\left[2
  \left(\frac{\ln(1-z)}{1-z}\right)_+  -\frac{\ln(z)}{1-z}- 
\frac{\ln(r^2)}{(1-z)_+}
\right]\right\}\,,
\eeeq
where $r=\mu/m_H$, and where the $+$ distribution is defined in the
usual way. Starting at NLO, there are also  
two new partonic channels, $qg\to H q$ and $q\bar{q}\to Hg$. Their
cross-sections are 
also known at ${\cal O}(\alpha_s)$
from~\cite{Djouadi:1991tka,Dawson:1990zj,Altarelli:1998gn}:  
\beeq
\omega^{qg\to H}\left(z,\alpha_s,r\right)&=&\frac{\alpha_s}{\pi}
\left\{-\frac{1}{3}(1-z)(7-3z)+\frac{2}{3}\left(1+(1-z)^2\right)
\left[\ln\frac{(1-z)^2}{zr^2} 
+1\right]\right\}\,,\nn\\[2mm]
\Delta \omega^{qg\to H}\left(z,\alpha_s,r\right)&=&\frac{\alpha_s}{\pi}
\left\{(1-z)^2+\frac{2}{3}\left(1-(1-z)^2\right) \left[\ln\frac{(1-z)^2}{zr^2}+1
\right]\right\}\,,
\eeeq
and 
\beq
\omega^{q\bar{q}\to
  H}\left(z,\alpha_s,r\right)\,=\,\frac{\alpha_s}{\pi}\,\frac{32}{27}\, 
(1-z)^3\,=\,-\Delta \omega^{q\bar{q}\to H}\,.
\eeq

We now compute the spin asymmetry $A_{\mathrm{LL}}$ in Higgs production at RHIC at
$\sqrt{S}=510$~GeV, as a function of the Higgs mass. We adopt the PDF
set with the positivity-violating negative 
gluon distribution of Ref.~\cite{Cocuzza:2022jye}, used already for
Fig.~\ref{fig1}.  
We also use the unpolarized PDFs of~\cite{Cocuzza:2022jye} for the
denominator of the spin asymmetry: this allows us to effectively
determine the individual helicity cross-sections in Eq.~(\ref{eq3})
and thus check positivity on a replica-by--replica basis. 
As a cross-check we have verified that for the LO and NLO unpolarized
cross-sections 
we recover the results given in Table~7 of
Ref.~\cite{Anastasiou:2016cez} when PDFs and parameters 
are chosen as in that paper. Figure~\ref{fig2} displays the asymmetry
$A_{\mathrm{LL}}$ for the replicas shown in Fig.~\ref{fig1}, 
on a linear (left) and on a logarithmic (right) scale, plotted as a
function of the Higgs mass.  For Higgs masses that deviate
from the physical value this should be taken as the result obtained in
a fictitious field theory in which the Yukawa couplings of quarks
are readjusted so that their masses remain the same and the
strongly-interacting sector of the theory remains the same.
A huge violation of the physical 
positivity condition~(\ref{eq2}) is observed. Already for Higgs masses
around the physical value 
the asymmetry $A_{\mathrm{LL}}$ exceeds unity; at even larger masses
it easily reaches values of 10 or even 100. Using instead the PDFs from
Ref.~\cite{Karpie:2023nyg}, which also include  lattice data,
 the positivity violation at the physical
Higgs mass value would likely be reduced
(see also Ref.~\cite{Hunt-Smith:2024khs}), but the
trend of Fig.~\ref{fig2}  suggests that it 
would  again be very large as the Higgs mass increases. We are thus led
to the conclusion that the 
PDF set with the positivity-violating negative gluon distribution
cannot be regarded as physical as  
it leads to negative cross-sections. 

It is important to note that the violation occurs in a kinematic
region corresponding to momentum fractions $x$ where the PDFs are
generally known best.  For
instance, at $m_h=150$~GeV and with $\sqrt{S}=510$~GeV we have
$x_1x_2\approx 0.09$ so at central rapidity $x_1=x_2\approx
0.3$. Hence,  the violation of physical positivity of
$A_{\mathrm{LL}}$ depends on the behavior of the PDF replicas in a
central $x$ region, and furthermore, it is clear from  
Fig.~\ref{fig2} that it is a bulk property the PDF replica
distribution.  Hence, it cannot be attributed to outliers, or to the
PDF behavior in very small  $x$ or large $x$ extrapolation regions: it
is not a consequence of statistical fluctuations or large uncertainties. 

Moreover, in this region unpolarized
PDFs, and even the gluon PDF, are known rather accurately. Indeed, the
uncertainty on the unpolarized gluon, taking the conservative
PDF4LHC21 combination, is about
5\%~\cite{PDF4LHCWorkingGroup:2022cjn}, so the violation of positivity
cannot be reasonably attributed to imperfect knowledge of the
unpolarized gluon and its uncertainty. Indeed, we have checked that
replacing the unpolarized PDFs of~\cite{Cocuzza:2022jye} in the
computation of the asymmetry with the PDFs of 
the PDF4LHC21 set~\cite{PDF4LHCWorkingGroup:2022cjn} and always taking
the largest of the 100  PDF4LHC21 replicas, which corresponds to a
more than three-$\sigma$ interval about the central gluon, we still
get positivity violation for $m_H\gtrsim140$~GeV, exponentially
increasing with Higgs mass with a pattern analogous to the curves
shown in Fig.~\ref{fig2}.

For comparison, we also show in Fig.~\ref{fig2} (lower bands) the
double-spin asymmetries obtained for the  sets
of~\cite{Cocuzza:2022jye} with {\it positive} $\Delta g$, which show a
strikingly different behavior. In this case, about 1000 PDF replicas
are available and plotted. The vast majority of them 
satisfies positivity. Violations of positivity are either in the tail
of the distribution, or in kinematic regimes dominated by very large
values of $x$. In these regions, the unpolarized gluon PDF is poorly
known, and it is in fact extrapolated from information at smaller $x$,
so the positivity violation could be reabsorbed in a change of
unpolarized gluon PDF. 

\begin{figure}[h!]
\vspace*{4mm}
\hspace*{-1.4cm}
\epsfig{figure=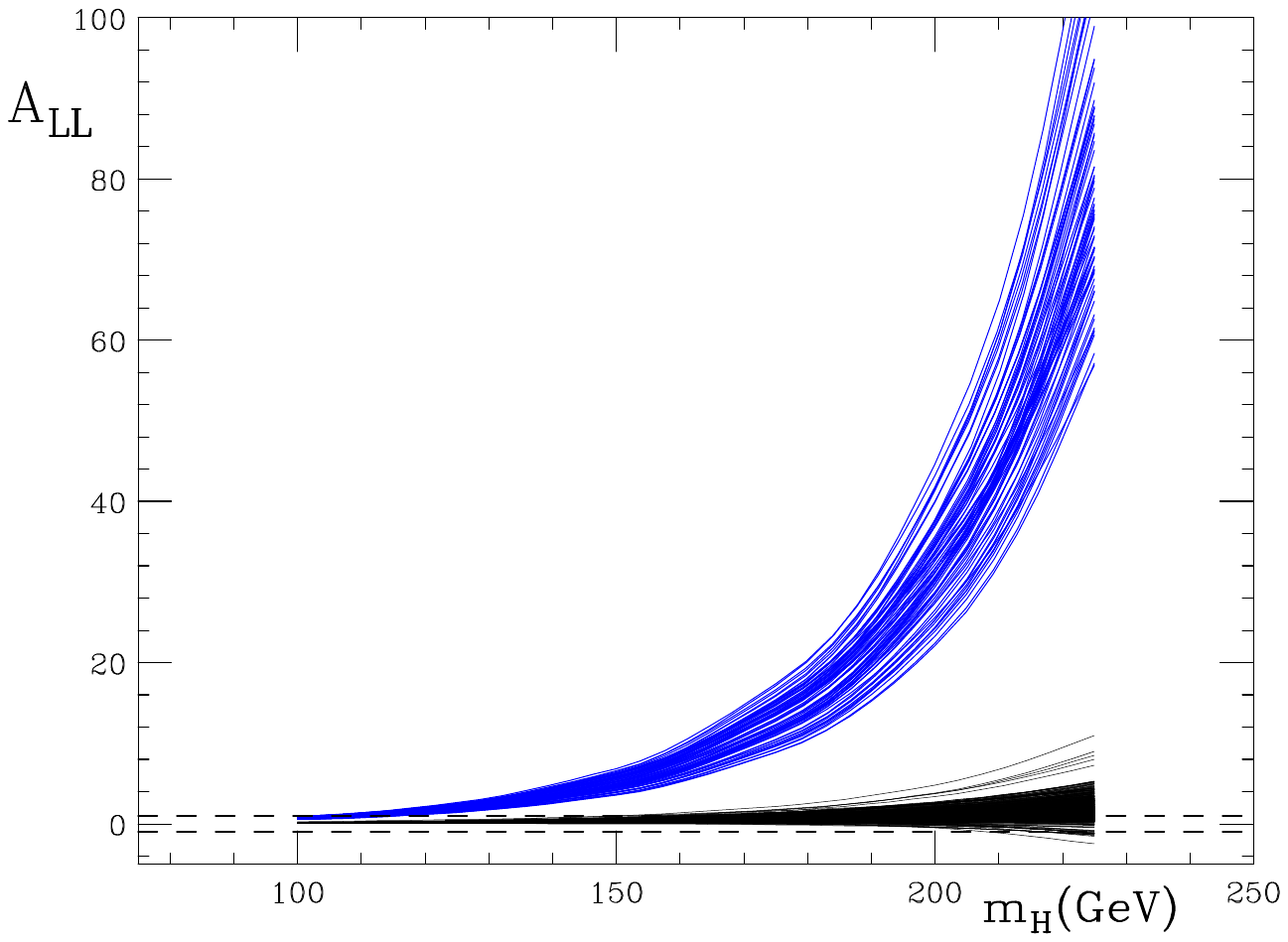,width=0.6\textwidth,clip=}
\hspace*{-1.6cm}
\epsfig{figure=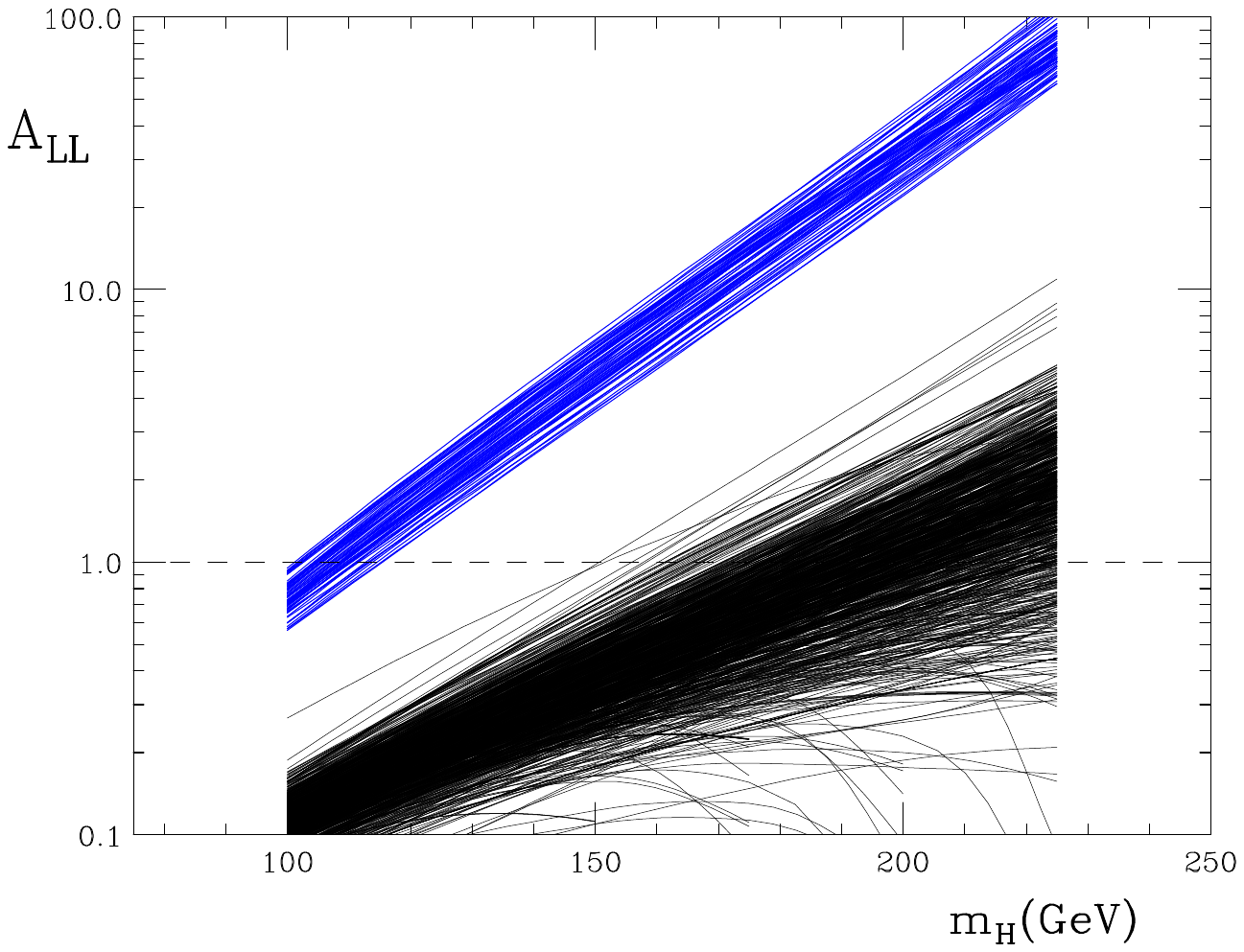,width=0.6\textwidth,clip=}
\vspace*{-1.3cm}
\caption{ \label{fig2} {\it Double-helicity asymmetry for Higgs production at
RHIC ($\sqrt{s}=510$~GeV) plotted as a function of the Higgs mass,
with a linear (left) or logarithmic (right) scale on the vertical
axis.  
The upper bands show $A_{{\mathrm{LL}}}$ 
as obtained for the gluon distribution shown in Fig.~\ref{fig1}, while
the lower bands provide the corresponding result for the sets
of~\cite{Cocuzza:2022jye} with $\Delta g\geq 0$. In both plots, the
dashed lines show the physical limit given by $|A_{\mathrm{LL}}|=1$.}} 
\end{figure}                 

As mentioned, in a consistent quantum field theory negative
cross-sections cannot occur for any process, regardless of whether it
is measurable in practice or even in principle. However, in this case
it is interesting to observe that it is a physically measurable
hadronic cross-section that is predicted to be negative. Indeed, the
gluon fusion  process dominates the Higgs production cross-section at
all energies (see
e.g.~\cite{Ellis:1996mzs,ParticleDataGroup:2022pth}). We have in fact
checked explicitly, using {\tt
  MadGraph5\_aMC@NLO}~\cite{Alwall:2014hca} that the vector-boson
fusion process, which dominates Higgs productions in the quark channel
$qq'\to qq' H$, despite the presence of initial-state valence quarks,
gives a contribution to Higgs production in proton-proton collisions
that is more than an order of magnitude smaller 
than that by gluon fusion
also at RHIC energy. Hence the positivity violation seen in
Fig.~\ref{fig2} leads to the prediction of a negative cross-section
for inclusive Higgs boson production in proton-proton collisions.  

Along the same lines, one may wonder whether, given the large size of
higher-order corrections to Higgs production in gluon fusion, the  
violation of physical positivity
seen in Fig.~\ref{fig2} could be due to these  NLO corrections, or
perhaps be alleviated by higher-order corrections. 
However, NLO corrections in fact cancel to a very large degree in the
spin asymmetry. Furthermore,  
the channels with incoming quarks, $qg\to H q$ and $q\bar{q}\to Hg$,
although nominally favored 
for very high $x$ thanks to the participation of a valence quark PDF,
remain sub-dominant and hence cannot re-instate positivity. It would
be straightforward to further improve the perturbative  
framework by carrying out threshold resummation for the Higgs
cross-section, following 
the lines of~\cite{Catani:2003zt}. This would, in fact, be required
for an accurate phenomenological  
study of Higgs production at RHIC. However, this, too, is irrelevant
for positivity since 
both the spin-averaged and the polarized cross-section receive the
same QCD corrections near 
partonic threshold $\hat{s}\approx m_H^2$. In fact, the latter
observation can be made  more 
general: the fact that the region of interest here is $x\gtrsim 0.25$
means for a $pp$ collider 
that any relevant process is probed close to partonic threshold. Given
that at threshold the QCD corrections  
are dominated by soft emission and that soft-gluon emission is
spin-independent, for such a kinematic regime the  
dominant QCD corrections will be very similar for general polarized
and unpolarized cross-sections, and even identical in some cases 
as for color-singlet $2\rightarrow 1$ Higgs production. Therefore,
spin asymmetries are given by their LO expression 
to high accuracy. Thus a significant violation of the LO positivity
condition such as Eq.~(\ref{eq1}) in a dominant partonic channel 
will automatically lead to the violation of positivity in the
physical hadronic cross-section and invalidate the corresponding
parton distributions.  

In conclusion,  by considering  Higgs boson production at RHIC we have 
shown that previously proposed scenarios for the proton's polarized
parton distribution functions 
with a large negative gluon polarization lead to unphysical negative
cross-sections. Reassuringly, such  
scenarios appear to be disfavored by RHIC data for
direct-photon~\cite{PHENIX:2022lgn}  
and dijet~\cite{RHICSPIN:2023zxx} production not included in the
analyses of Refs.~\cite{Zhou:2022wzm,Cocuzza:2022jye,Karpie:2023nyg},
as well as by the currently most advanced lattice study of $\Delta g$~\cite{Khan:2022vot}. 
Amusingly, the  Higgs production process that we have considered is in
fact not hypothetical at all:  based on our results,  
we estimate that about half a dozen Higgs bosons should have been
produced at RHIC 
during its lifetime with $510$~GeV running.

\section*{Acknowledgments}
This work originated from discussions at the Center for Frontiers in
Nuclear Science (CFNS) workshop ``Precision QCD predictions for ep
Physics at the EIC (II)'', Stony Brook, September 18-22, 2023. We are
grateful to M. Zaro for helpful remarks on the VBF process, and to
I. Borsa, J. Karpie, R. Sassot, N. Sato and M. Stratmann for
discussions and to N. Sato for comments on a first draft. The work of
D.de F. has been partially  
supported by Conicet, ANPCyT and the Alexander von Humboldt Foundation. W.V. is supported by Deutsche Forschungsgemeinschaft (DFG) through the Research Unit FOR 2926 (project 409651613).

\end{document}